\documentstyle[preprint,aps]{revtex}

\tightenlines

\begin{document}

\draft

\title{\bf An Application of Feynman-Kleinert Approximants to the Massive Schwinger Model on a Lattice}
\author{T.M.R. Byrnes, C.J. Hamer, Zheng Weihong and S. Morrison}
\address{School of Physics,                                              
The University of New South Wales,                                   
Sydney, NSW 2052, Australia.}                      

\date{\today}

\maketitle 

\begin{abstract}
A trial application of the method of Feynman-Kleinert approximants is
made to perturbation series arising in connection with the lattice
Schwinger model. In extrapolating the lattice strong-coupling series to
the weak-coupling continuum limit, the approximants do not converge
well. In interpolating between the continuum perturbation series at
large fermion mass and small fermion mass, however, the approximants do
give good results. In the course of the calculations, we picked up and
rectified an error in an earlier derivation of the continuum series
coefficients.

\end{abstract}                    

\narrowtext

\section{INTRODUCTION}

The analytic continuation of a function, whose perturbation series expansion 
about a given point is known, is an important and long-standing problem. Some 
standard methods are the use of Pad\'{e} approximants, or integrated 
differential approximants, as reviewed for instance by 
Guttmann\cite{gut89}. In some circumstances, these techniques can work 
extremely well; but in others they fail. A well known difficult case is that 
of a function whose series expansion about the origin in a variable $ x $, say,
 is known up to a given order, and we want to extrapolate to 
$ x \rightarrow \infty $. If there is a cut or an essential singularity at 
infinity, the standard methods do not work well. This is a common situation in 
lattice gauge theory, for instance. 

In recent years, a new method has been developed, involving ``Feynman-Kleinert 
approximants''\cite{kle95}. This is a variational technique, arising originally
 from a Gaussian approximation to the path integral\cite{fey86}. It has been used 
with great success on a number of problems, such as extrapolation of the 
weak-coupling series for the anharmonic oscillator\cite{jan95}, where the asymptotic
 strong-coupling behaviour was obtained with extraordinary accuracy, up to 20 
significant digits. For a discussion and further references, see the 
recent paper by Kleinert\cite{kle98}. 

Here we attempt to apply the new method to a lattice gauge theory, namely the 
Schwinger model in (1+1) dimensions. In the series approach to lattice gauge 
theory, one generates strong-coupling series expansions for the physical 
quantities of interest, and then tries to extrapolate to the continuum, 
weak-coupling limit. It would be an important breakthrough if a better 
technique could be found for performing this extrapolation. In the report we 
set out to test whether Feynman-Kleinert approximants can fulfill this need. 
Unfortunately, we find that the technique is not very successful in this context. 

Another common problem is that of interpolation of a function, where a number of series coefficients are known of {\it both} the weak-coupling and the strong-coupling expansions. An example of this arises for the low-lying energy eigenvalues in the continuum Schwinger model, as functions of $ m/g $, where $ m $ is the fermion mass and $ g $ is the electric coupling. We show that Feynman-Kleinert approximants can provide an effective interpolation between these two limits.

\section{THEORY}
\label{sec2}

For reference, let us briefly recapitulate the algorithm set out 
by Kleinert\cite{kle95} using his notation. Consider a quantity $ E $ whose 
perturbation series expansion in some parameter $ \alpha $ is known to order 
$ N $:
\begin{equation}
E_N = \sum_{n=0}^N a_n \alpha^n .
\end{equation}
Write
\begin{equation}
\label{eqn2}
E_N^\omega = \omega^p \sum_{n=0}^N a_n \left( \frac{\alpha}{\omega^q} \right)^n,
\end{equation}
where $ \omega $ is an auxiliary parameter whose value will eventually be set equal to 1, and $ p,q $ are indices which will be chosen later to suit the problem at hand. Now set $ \omega \equiv \sqrt{\Omega^2 + \omega^2 - \Omega^2} $, and re-expand $ E_N^\omega $ in powers of $ \lambda = 1 - \omega^2/\Omega^2 $, treating $ \lambda $ as a quantity of order $ \alpha $, and truncating the re-expanded series at order $ N $. The result is a function
\begin{equation}
W_N (\alpha, \Omega) = \Omega^p \sum_{n=0}^N a_n f_n (\Omega) \left( \frac{\alpha}{\Omega^q} \right)^n ,
\end{equation}
where
\begin{equation}
f_n (\Omega) = \sum_{j=0}^{N-n} \left(
\begin{array}{c}
(p-qn)/2 \\
j
\end{array}
\right) (-1)^j \left( 1- \frac{\omega^2}{\Omega^2} \right)^j .
\end{equation}
Forming the first and second derivatives of $ W_N (\alpha, \Omega) $ with respect to $ \Omega $ (and setting $ \omega=1 $), we find the positions of the turning points. The smallest among these is denoted $ \Omega_N $. The resulting $ W_N (\alpha ) \equiv W_N (\alpha, \Omega_N ) $ gives the desired approximation to the function $ E $ at finite $ \alpha $, an {\it extrapolation} of the function $ E_N $ to finite $ \alpha $. 

Now consider the limit $ \alpha \rightarrow \infty $. At large $ \alpha $, $ \Omega_N $ will scale with $ \alpha $ like 
\begin{equation}
\label{eqn5a}
\Omega_N \sim \alpha^{1/q} c, 
\end{equation}
so that
\begin{equation}
W_N (\alpha, \Omega_N) \sim \alpha^{p/q}.
\end{equation}
The full `strong-coupling' (large $ \alpha $) expression can now be obtained by writing
\begin{equation}
W_N (\alpha, \Omega) = \Omega^p w_N (\hat{\alpha}, \omega^2/\Omega^2 ) ,
\end{equation}
with $ \hat{\alpha} = \alpha/ \Omega^q $, and expanding $ w_N $ in powers of 
$ \omega^2/\Omega^2 $, which for $ \alpha \rightarrow \infty$ behaves
like $(1/c^2)(\alpha/\omega^q)^{-2/q}$. The result is
\begin{equation}
\label{eqn7}
W_n (\alpha) = \alpha^{p/q} \left[ b_0 (c) + b_1 (c) \left( \frac{\alpha}{\omega^q} \right)^{-2/q} + b_2 (c) \left( \frac{\alpha}{\omega^q} \right)^{-4/q} + \dots \right]
\end{equation}
with
\begin{eqnarray}
\nonumber
b_n (c) & = & \frac{1}{n!} w_N^{(n)} ( \hat{\alpha}, 0) \left. 
\hat{\alpha}^{(2n-p)/q} \right|_{\hat{\alpha} = 1/c^q}\\
& = & \sum_{l=0}^N a_l \sum_{j=n}^{N-l} \left(
\begin{array}{c}
(p-lq)/2 \\
\label{eqn8}
j
\end{array} \right)
\left(
\begin{array}{c}
j \\
n
\end{array}
\right)
(-1)^{j-n} c^{p-lq-2n} .
\end{eqnarray}
The indices $ p $ and $ q $ in the expansion (\ref{eqn2}) are now chosen so as to give the correct asymptotic behaviour (i.e. the correct powers of $ \alpha $) in the strong-coupling expansion (\ref{eqn7}), which we assume is known {\it a priori}. The leading coefficient $ c $ in (\ref{eqn5a}) is found by searching for the extrema of the leading coefficient $ b_0 (c) $ as a function of $ c $ and choosing the smallest of them. 

Finally, we have to account for the fact that $ \Omega_N $ will have corrections to the leading behaviour $ \alpha^{1/q} c $.  We can allow for this by replacing $ c $ by
\begin{equation}
c (\alpha) = c_0 + c_1 \left( \frac{\alpha}{\omega^q} \right)^{-2/q} + c_2 \left( \frac{\alpha}{\omega^q} \right)^{-4/q} + \dots
\end{equation}
requiring a re-expansion of the coefficients $ b_n (c) $ in (\ref{eqn8}). The expansion coefficients $ c_n $ are determined by looking for the turning points of $ b_{2n} (c)$, successively. 

The equations
\begin{equation}
\label{eqn11}
\frac{ d b_{2n} (c)}{d c_n} = 0 ,
\end{equation}
together with knowledge of the coefficients $ \{ a_n, n=0,  \dots , N \} $, 
can in principle be used to determine a set of parameters $ \{ c_n, n = 0, 
\dots, N \} $, and hence estimate the leading strong-coupling coefficients 
$ b_n (c) $. This provides estimates of the {\it asymptotic behaviour} of the 
function $ E $. In the case of the strong-coupling anharmonic oscillator, it 
has been shown\cite{jan95} that the resulting estimates converge exponentially 
fast, and have been used to estimate the energy eigenvalue in the 
strong-coupling limit accurate to 20 decimal places!

Alternatively, any knowledge of the strong-coupling coefficients $ b_0, b_1, \dots $ can be fed back into the set of equations (\ref{eqn8}) and (\ref{eqn11}) to obtain estimates of further weak-coupling coefficients $ a_{N+1}, a_{N+2}, \dots $, and thus carry the extrapolation $ W_N (\alpha ) $ to higher orders. This is the basis of the {\it interpolation} algorithm between weak and strong-coupling for the function $ E $.

\section{RESULTS}

\subsection{Extrapolation to the Continuum Limit}

We have applied the Feynman-Kleinert approximation method to the lattice 
Schwinger model, as formulated by Banks {\it et al.} \cite{banks76}. The 
rescaled lattice Hamiltonian is 
\begin{equation}
W = \frac{2}{a g^2} H = W_0 + x V ,
\end{equation}
where
\begin{eqnarray}
W_0 = \sum_n L^2 (n) + \mu \sum_n (-1)^n \phi^\dagger (n) \phi (n), \\
V = -i \sum_n \left[ \phi^\dagger (n) e^{i \theta (n)} \phi(n+1) - \mbox{h.c.} \right],
\end{eqnarray}
and
\begin{equation}
\mu = \frac{2m}{g^2 a}, \hspace{1cm} x = \frac{1}{g^2 a^2} .
\end{equation}
Here $ m $ is the electron mass, $ g $ is the electric coupling constant, and $ a $ is the lattice spacing. The field $ \phi(n) $ is the lattice fermion field defined on sites $ n $, 
\begin{equation}
\{ \phi (n), \phi^\dagger (m) \} = \delta_{nm} ,
\end{equation}
while $L(n)$ is the electric field on link $n$, which takes only integer
eigenvalues.

Expansions of the energy eigenvalues of the operator $ W $ have been calculated
 to high orders in the perturbation parameter $ x $, using efficient linked 
cluster methods\cite{ham97}. Our object now is to see whether these series can 
be accurately extrapolated to the continuum limit, which corresponds to $ a 
\rightarrow 0 $ or $ x \rightarrow \infty $, using the Feynman-Kleinert approach.

\subsubsection{Ground State Energy}

For the ground state energy per site $ \omega_0/M $, where $ M $ is the number of states, the series expansion is of the form
\begin{equation}
\omega_0/M = \sum_n a_n \left( x^2 \right)^n ,
\end{equation}
where $ a_0 = 0 $; and the asymptotic behaviour as $ x \rightarrow \infty $ is 
known\cite{banks76}
\begin{equation}
\label{eqn19}
\omega_0/M \sim - \frac{2 x}{\pi} .
\end{equation}
The coefficients $ a_n $ have been calculated up to order $ n = 15 $ 
\cite{ham97}. In this case, we can apply the formalism of section \ref{sec2}, taking $ E = \omega_0/M $, $ \alpha = x^2 $, $ p=2 $, $ q=4 $, which implies the asymptotic behaviour as $ x \rightarrow \infty $
\begin{equation}
\omega_0/M \sim b_0 x + b_1 + O ( x^{-1} ) .
\end{equation}

\begin{figure}
\centerline{\psfig{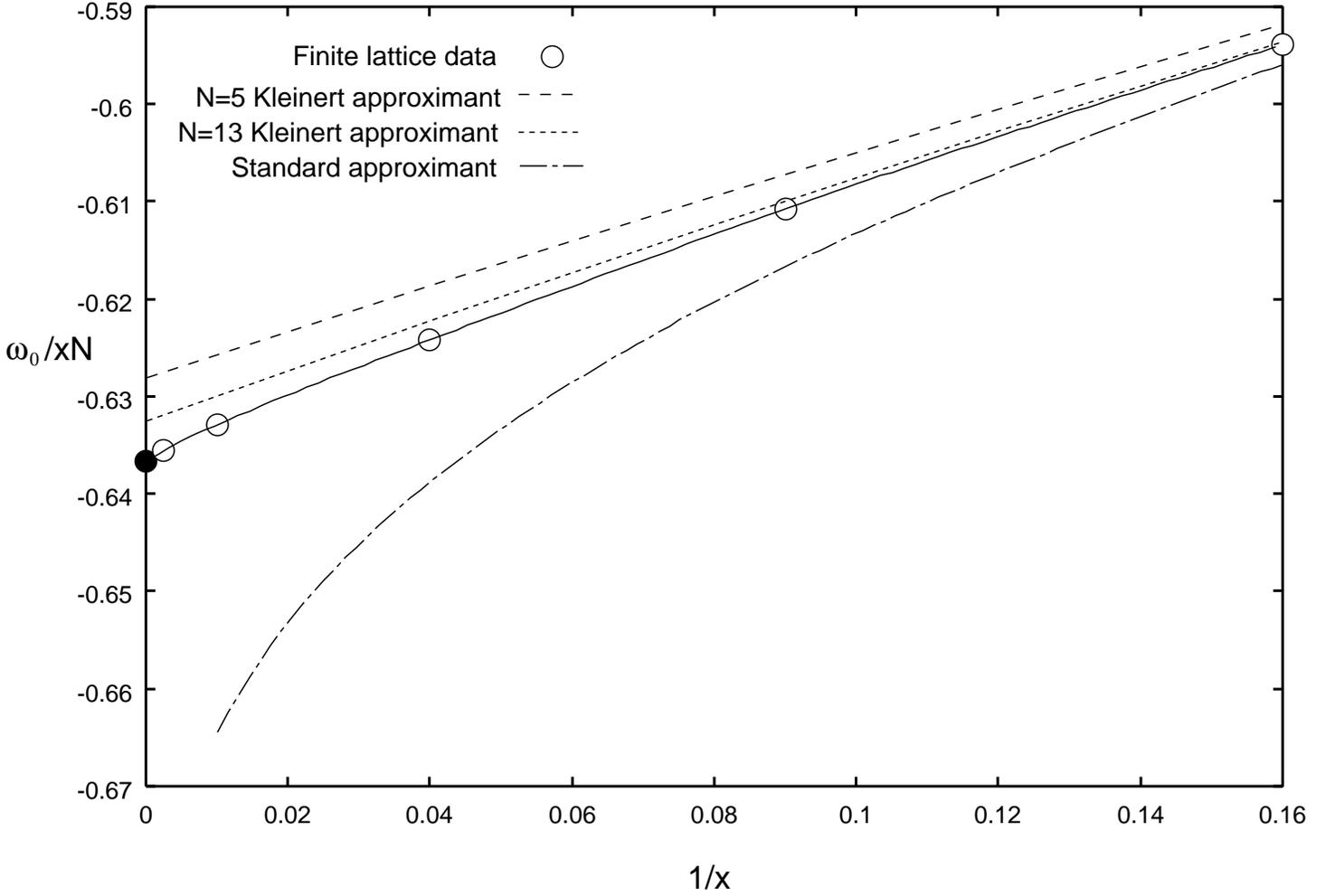}}
\caption{Ground state energy per site of the lattice Schwinger model at $ m/g = 0 $, as obtained by the Kleinert approximants with $ N =5 $ and $ N = 13 $. Open circles denote finite lattice calculations, the closed circle shows the position of the exact result in the continuum limit, $ \omega_0/xN = -2/\pi $. We also show an extrapolation using a standard Pad{\'e} approximant for comparison. }
\label{fig1}
\end{figure}

Fig. \ref{fig1} shows the resulting approximants to the ground-state energy at $ m/g = 0 $ compared with numerical results obtained by other methods, and the known continuum limit, Eq. (\ref{eqn19}). It can be seen that the approximants do seem to converge to the correct limit, but only rather slowly.

\subsubsection{Energy Gap}

\begin{figure}
\centerline{\psfig{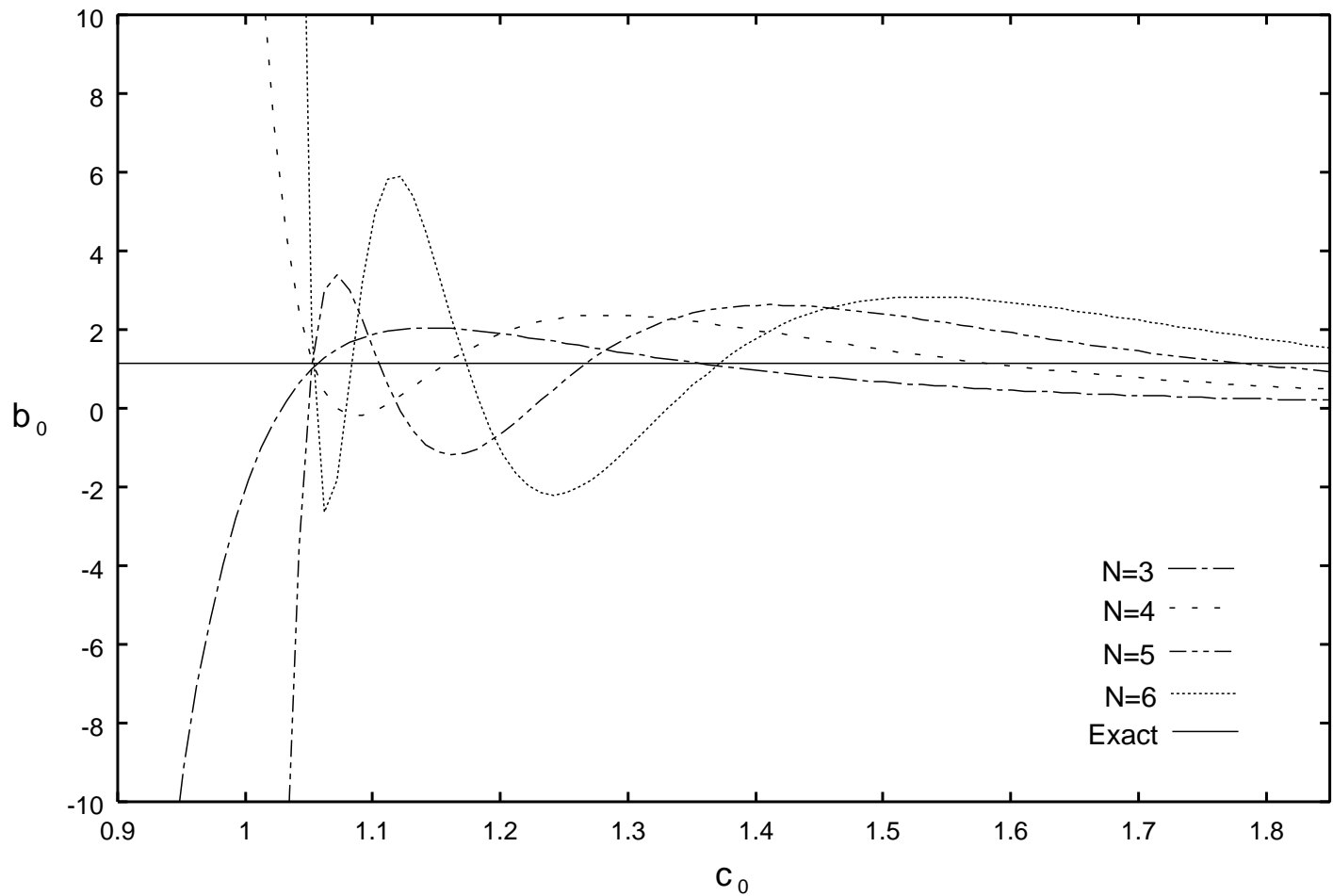}}
\caption{Dependence of the value $ b_0 $ as a function of $ c_0 $ for the case $ m/g = 0 $, at four orders in $ N $. The exact result is shown as a solid line at $ b_0 = 2/\sqrt{\pi} \approx 1.128 $. }
\label{fig2}
\end{figure}

The energy gap to the lowest-lying positronium state has an expansion
\begin{equation}
\omega_1 - \omega_0 = \sum_n a_n \left( x^2 \right)^n
\end{equation}
and the expected asymptotic behaviour as $ x \rightarrow \infty $ is
\begin{equation}
\omega_1 - \omega_0 \sim b_0 x^{1/2} .
\end{equation}
The coefficients $ a_n $ for this series have been calculated up to order 
$ n= 14 $ \cite{ham97}. Here, we take $ \alpha = x^2 $, $ p=2 $, $ q=8 $, 
assuming the asymptotic behaviour as $ x \rightarrow \infty $
\begin{equation}
\omega_1 - \omega_0 \sim b_0 x^{1/2} + b_1 + O (x^{-1/2})
\end{equation}
which accords with previous studies\cite{ham97,byrnes02}. In this case, however, 
the Feynman-Kleinert approximants do not converge at all at large $ x $. Fig.
 \ref{fig2} shows the dependence of the estimated value of $ b_0 $ as a 
function of the parameter $ c_0 $, for the case $ m/g = 0 $, at various orders 
$ N $. It can be seen that as $ c_0 $ increases, the estimate of $ b_0 $ comes 
in from infinity, oscillates around the correct value (marked by a solid
line
), and then slowly drifts away again. Unfortunately, the oscillations increase 
with order $ N $, and the values at the lowest turning point correspondingly 
diverge away from the correct value as $ N $ increases. The same is true of 
the values at the lowest point of inflection. It seems that the Feynman-Kleinert approximant method fails in this case.

\subsection{Interpolation of Continuum Series}

Another problem connected with series expansions for the Schwinger model arises
 in connection with the continuum field theory, rather than the lattice model. 
A few coefficients are known of the series expansions for the lowest-lying 
excited states of the model, the ``vector'' and ``scalar'' positronium states, 
in both the strong-coupling and weak-coupling regimes. At strong coupling, 
$ m/g \rightarrow 0 $, the first three coefficients have been computed by 
Adam\cite{adam}, using the bosonic form of the theory:
\begin{equation}
\frac{E}{g} = \left( \frac{M-2m}{g} \right) = \sum_n a_n \left( m/g \right)^n .
\end{equation}
At weak coupling, $ m/g \rightarrow \infty $, the first three coefficients have been calculated by Sriganesh {\it et al.} \cite{sriganesh00}, following a treatment of Coleman \cite{coleman76}:
\begin{equation}
\frac{E}{g} \sim \left( \frac{g}{m} \right)^{1/3} \sum_n b_n\left[ \left( \frac{g}{m} \right)^{2/3} \right]^n .
\end{equation}

The problem now is, how can one compute a sensible interpolation between these two series to obtain accurate estimates of the energy eigenvalues at finite $ m/g $? One possibility is to use ``2-point'' Pad\'{e} approximants: but they involve only integer powers of the variable, $ (m/g) $ in this case. By fitting an appropriate power of the function $ E/g $ one can mimic the leading behaviour as $ m/g \rightarrow \infty $, but cannot reproduce the sub-leading powers. Furthermore, experience shows that 2-point Pad\'e approximants are prone to exhibit spurious `wiggles' at intermediate couplings between the two limits. Here we set out to apply Feynman-Kleinert approximants to the problem.

Setting $ \alpha = m/g $, $ p=1 $, $ q=3 $, we have computed Feynman-Kleinert 
approximants for the energy eigenvalues using known values for the coefficients
 $ \{ a_n, b_n, n=0,N-1 \} $ for $ N = 1,2,3 $. Now knowing $ 2N $ coefficients
 in total allows us to estimate the next $ N $ coefficients for either the 
$ a_n $ or $ b_n $ series. Table \ref{tab1a} shows the `predicted' values 
versus the known values for $ N = 1 $. For the vector state with $ N=1 $, the 
predicted values are in quite good agreement with the known values, to within a
 few percent. 
At $ N= 2$ (see Table \ref{tab1b}), however, a large discrepancy becomes 
evident between the predicted and `known' values for the coefficient $ b_2 $ - 
even the sign is different. This prompted us to re-examine the calculation of the weak-coupling 
coefficients $ b_n $ by Sriganesh {\it et al.}\cite{sriganesh00}; and indeed we discovered some 
errors in those calculations. A corrected version of the calculations is given 
in the Appendix; the corrected series coefficients are also shown in Table 
\ref{tab1b}. 

For the scalar state, the `predicted' series coefficients disagree with
the known ones even at $ N = 1 $ - but this is easily understood, in that
the simplest interpolation between the zeroth order series at either end
would have no `hump' in the middle such as is shown in Fig. \ref{fig4}.
The predicted values at $ N = 2 $ agree qualitatively with the (corrected)
known values.

\begin{figure}
\centerline{\psfig{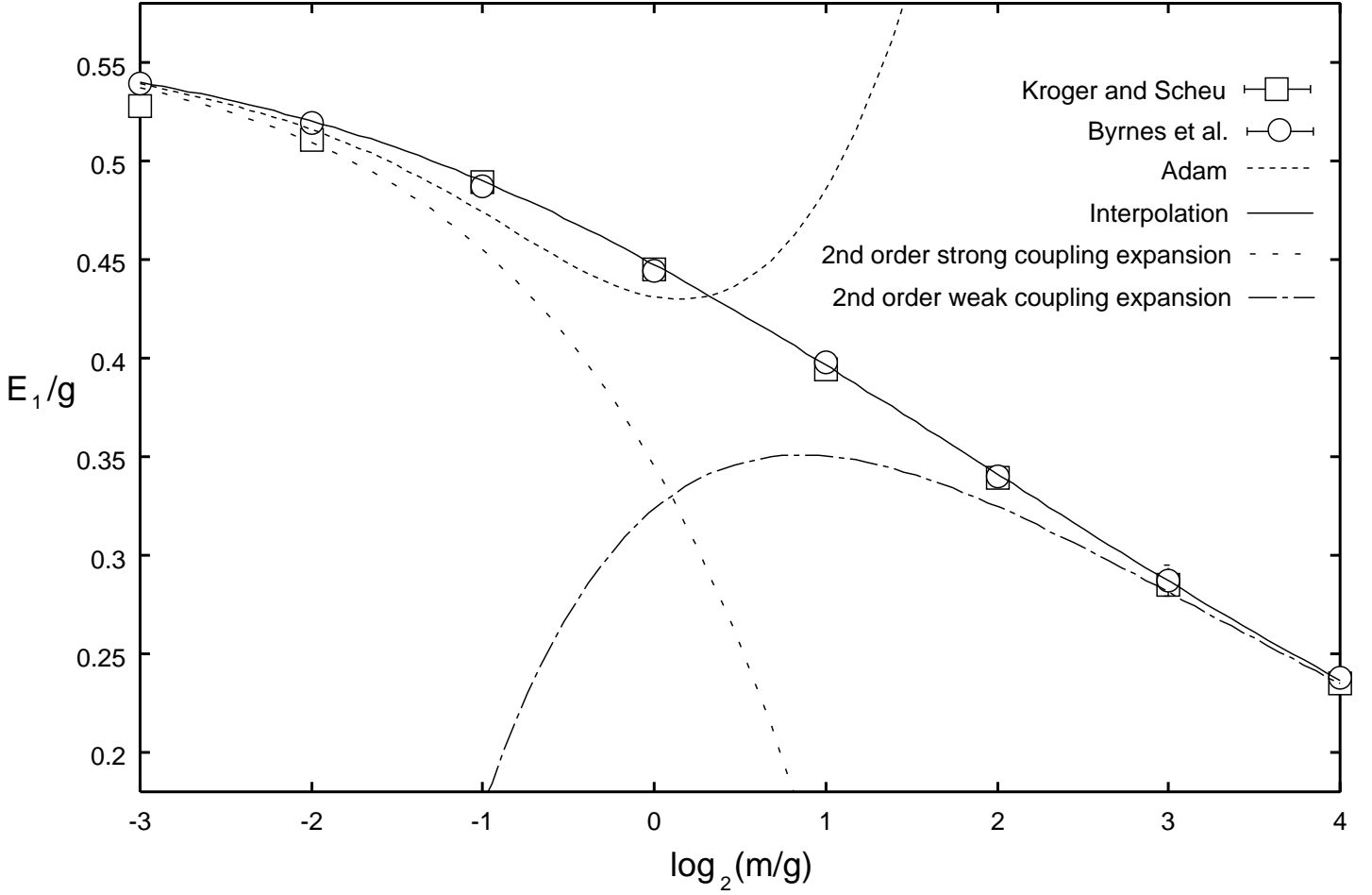}}
\caption{A comparison of the Feynman-Kleinert interpolation for the vector state with numerical DMRG data of Byrnes {\it et al.} \protect\cite{byrnes02}, ``fast moving frame'' estimates of Kr{\"o}ger and Scheu \protect\cite{kroger98}, and the ``renormal-ordered'' mass perturbation theory results of Adam \protect\cite{adam02}. The 2nd order ($ N=2 $) weak and strong coupling series, which are used to generate the interpolation, are also shown for comparison.}
\label{fig3}
\end{figure}

\begin{figure}
\centerline{\psfig{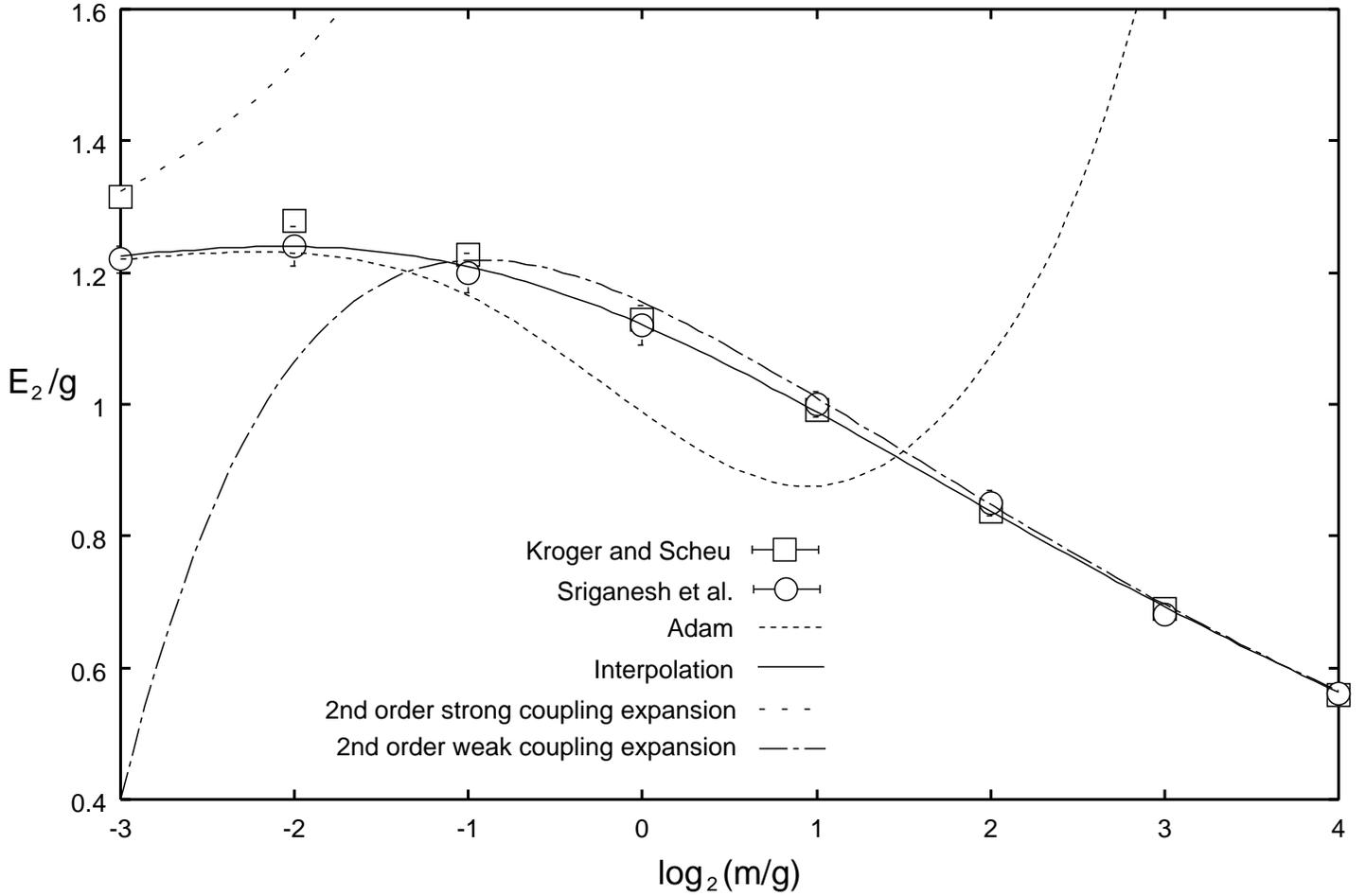}}
\caption{A comparison of the Feynman-Kleinert interpolation for the scalar state to finite lattice estimates of Sriganesh {\it et al.} \protect\cite{sriganesh00}, ``fast moving frame'' estimates of Kr{\"o}ger and Scheu \protect\cite{kroger98}, and the ``renormal-ordered'' mass perturbation theory results of Adam \protect\cite{adam02}. The 2nd order ($ N=2 $) weak and strong coupling series, which are used to generate the interpolation, are also shown for comparison.}
\label{fig4}
\end{figure}

The Feynman-Kleinert approximants were found to give smooth and apparently 
accurate interpolations of the energy eigenvalues at finite $ m/g $, up to 
order $ N=2 $. At order $ N=3 $, no solution was found to the system of 
simultaneous equations - we have not explored in detail why this occurred. 
Figures \ref{fig3} and \ref{fig4} show the results for the vector and scalar 
state interpolations. For the vector state energies, our interpolation is 
compared to data obtained from the density matrix renormalization group (DMRG) 
method \cite{byrnes02}, the ``fast moving frame'' estimates of Kr{\"o}ger and Scheu \cite{kroger98}, and the ``renormal-ordered'' mass perturbation series results of Adam \cite{adam02}. The scalar state interpolations are compared to the finite lattice calculations
of Sriganesh {\it et al.} \cite{sriganesh00}, and again results of Kr{\"o}ger and Scheu,  and Adam. Both plots show excellent agreement between
 the interpolation and the numerical data, within errors. 
We also plot the second order strong and weak coupling series used to generate the interpolation on the same plots, which diverge away for large and small masses respectively. The ``renormal-ordered'' expansion \cite{adam02} about $ m/g = 0 $, extends the region of validity of the perturbation theory, and agrees fairly well with our interpolation for small masses. 

The interpolation performs particularly well for the scalar state, where 
for small $ m/g $, the original strong coupling series disagrees significantly 
with the numerical data. The interpolation on the other hand agrees extremely well with existing results, despite the large gap that must be bridged between the two 
series. Tables \ref{tab3a} and \ref{tab3b} list numerical values for specific values of $ m/g $ for the interpolation compared to previous works. 
We estimate the error on our interpolations based on the difference between our quoted $ N = 2 $ calculation and the corresponding $ N = 1 $ calculation. 
Error estimates have probably been overestimated somewhat using this method,
as can be seen from the good agreement with the numerical data in Figure \ref{fig4}. We have not however found a better way of estimating these errors.

\section{CONCLUSIONS}

The Feynman-Kleinert approximants have earned a mixed scorecard in the 
applications we have discussed. For the extrapolation of lattice 
strong-coupling series to the weak-coupling continuum limit, the approximants 
have basically failed. They converged for the ground-state eigenvalue, but 
only slowly; and for the first excited state eigenvalue, they did not converge 
at all. This is probably due to the presence of exponentially decaying terms 
corresponding to essential singularities in the weak-coupling limit, which are 
not accounted for in the Feynman-Kleinert approach. The upshot is, 
unfortunately, that this approach does not seem to have any advantage over 
previous techniques\cite{ham97}, which used Pad\'e or integral approximants to 
extrapolate to intermediate couplings, and then `matched' the results onto a 
weak coupling form to reach the continuum limit. 

For the continuum field theory, however, the Feynman-Kleinert approximants have
 proved useful in providing a sensible and apparently accurate interpolation 
between the known weak coupling and strong coupling series. The fact that we 
were able to pick up a mistake in the series coefficients by this approach 
adds confidence in the technique. No doubt many applications of a similar 
nature are possible.

\begin{acknowledgements}
We are grateful to Dr. C. Adam for very useful correspondence on this topic. 
This work forms part of a research project supported by the Australian
Research Council.
\end{acknowledgements}

\appendix

\section{WEAK COUPLING EXPANSION OF THE MASSIVE SCHWINGER MODEL}

Our starting point is the Hamiltonian as derived by Coleman in the two-particle subspace (see Eq. (4.18) of Ref. \cite{coleman76}). We set the background field $ \theta = 0 $. Expanding this Hamiltonian for large mass $ m $, and keeping terms to order $ O(m^{-3}) $ we obtain a Schr\"odinger equation
\begin{equation}
(H_0 + H_1) \psi(x) = E \psi(x) ,
\label{eq25}
\end{equation}
where
\begin{eqnarray}
\label{expandedham}
H_0 & = &  - \frac{1}{m} \frac{\partial^2}{\partial x^2} + \frac{g^2}{2} |x| \\
H_1 & = &  - \frac{g^2}{ m \pi} - \frac{1}{4m^3} \frac{\partial^4}{\partial x^4} + \frac{g^2}{4m^2} \delta(x) - \frac{g^2}{2 \pi m^3} \frac{\partial^2}{\partial x^2}.
\end{eqnarray}
Rescale to dimensionless variables as follows:
\begin{equation}
\label{changeofvariables}
x = \left( \frac{2}{m g^2} \right)^{1/3} z \hspace{1cm}
E = \left( \frac{g^4}{4m} \right)^{1/3} \epsilon .
\end{equation}
Then (\ref{eq25}) becomes 
\begin{equation}
(H'_0 + H'_1) \psi(z) = \epsilon \psi(z) ,
\end{equation}
where
\begin{eqnarray}
H'_0 & = &  -  \frac{\partial^2}{\partial z^2} +  |z| \\
H'_1 & = &  -\frac{1}{\pi} \left(\frac{2g}{m} \right)^{2/3} - \left( \frac{g}{4m} \right)^{4/3} [ 
\frac{\partial^4}{\partial z^4} - 2 \delta(z) ] .
\end{eqnarray}
The last term in (\ref{expandedham}) has now been dropped as it is now clear it is of order $ \sim (g/m)^2 $. The solution of the leading-order Schr{\" o}dinger equation
\begin{equation}
\label{1storderschrodinger}
\left( \frac{d^2}{d z^2} - |z| \right) \psi(z) = -\epsilon \psi(z) ,
\end{equation}
was discussed by Hamer\cite{hamer77}. The solution of
(\ref{1storderschrodinger}) is a
symmetric or antisymmetric Airy function which obeys the condition
\begin{eqnarray}
\mbox{Ai}'(-\epsilon_n) = 0 & & \hspace{1cm} \mbox{(symmetric)} \\
\mbox{Ai}(-\epsilon_n) = 0 & &  \hspace{1cm} \mbox{(antisymmetric)} .
\end{eqnarray}
The lowest symmetric (antisymmetric) wavefunction gives the energy for the 
vector (scalar) state.

 The higher order terms contained in $ H'_1 $ may be taken into account by 
calculating the correction term $ \Delta \epsilon = \langle \psi | H'_1 | \psi \rangle $ using numerical integration (the results are shown in Table \ref{tab4}). Rescaling back to our original variables gives us
\begin{equation}
\frac{E_1}{g} = 0.6418 \left( \frac{g}{m} \right)^{1/3} - \frac{1}{\pi} \left( \frac{g}{m} \right)
+ 0.1547 \left( \frac{g}{m} \right)^{5/3} ,
\end{equation}
for the vector state binding energy, and
\begin{equation}
\frac{E_2}{g} = 1.473 \left( \frac{g}{m} \right)^{1/3} - \frac{1}{\pi} \left( \frac{g}{m} \right)
- 0.1093 \left( \frac{g}{m} \right)^{5/3} .
\end{equation}
for the scalar state binding energy.

\setdec 0.00000000000
\begin{table}
\caption{Comparison of known coefficients in weak \protect\cite{sriganesh00} and strong \protect\cite{adam} coupling series, versus the Feynman-Kleinert estimates (labeled as ``Predicted''). Feynman-Kleinert estimates are obtained by starting with zeroth order coefficients $ a_0 $ and $ b_0 $ (i.e. $ N= 1 $), for each of the vector and scalar particle states.
\label{tab1a} }
\begin{tabular}{ccccc}
Coefficient & \multicolumn{2}{c}{Vector State} & \multicolumn{2}{c}{Scalar State} \\
      & Predicted & Known & Predicted & Known \\
\hline
$ a_1 $ & -0.2046 & -0.2189 & -0.2709 & 1.562 \\
$ b_1 $ & -0.2917 & -0.3183 & -0.8813 & -0.3183 \\
\end{tabular}
\end{table}
\begin{table}
\caption{As for Table \ref{tab1a}, but with $ N=2 $. The ``known'' coefficients for $ b_2 $ contain both the result obtained in Ref. \protect\cite{sriganesh00} (in square brackets), and our recalculation (unbracketed). 
\label{tab1b} }
\begin{tabular}{ccccc}
Coefficient & \multicolumn{2}{c}{Vector State} & \multicolumn{2}{c}{Scalar State} \\
      & Predicted & Known & Predicted & Known \\
\hline
$ a_2 $ & 0.2298 & 0.1907 & -10.13 & -13.51 \\
$ a_3 $ & -0.3148 &  & 49.01 &  \\
$ b_2 $ & 0.2159 & 0.1547 [-0.2521] & -0.1854 & -0.1093 [0.1085] \\
$ b_3 $ & -0.1669 &  & 0.2794 &  \\
\end{tabular}
\end{table}
\begin{table}
\caption{Comparison of interpolation results for the vector state with numerical results obtained by Byrnes {\it et al.} \protect\cite{byrnes02}, Kr{\"o}ger and Scheu \protect\cite{kroger98}, and Adam \protect\cite{adam02}. 
\label{tab3a}}
\begin{tabular}{ccccc}
$ m/g $ & This work  & Byrnes  & Kr{\"o}ger and & Adam \cite{adam02} \\
	&   	     & {\it et al.} \cite{byrnes02} & Scheu \cite{kroger98} & \\
\hline
0.125 & 0.540(2) & 0.53950(7) & 0.528 & 0.539	\\
0.25  & 0.520(3) & 0.51918(5) & 0.511 & 0.516	\\
0.5   & 0.490(4) & 0.48747(2) & 0.489 & 0.474	\\
1     & 0.448(4) & 0.4444(1)  & 0.455 & 0.43	\\
2     & 0.396(4) & 0.398(1)   & 0.394 & 0.49	\\
4     & 0.341(3) & 0.340(1)   & 0.339 & 0.76	\\
8     & 0.287(2) & 0.287(8)   & 0.285 & 1.40	\\
16    & 0.237(2) & 0.238(5)   & 0.235 & 2.75	\\
\end{tabular}
\end{table}
\begin{table}
\caption{Comparison of interpolation results for the scalar state with numerical results obtained by Sriganesh {\it et al.} \protect\cite{sriganesh00}, Kr{\"o}ger and Scheu \protect\cite{kroger98}, and Adam \protect\cite{adam02}. 
\label{tab3b}}
\begin{tabular}{ccccc}
$ m/g $ & This work  & Sriganesh & Kr{\"o}ger and & Adam \cite{adam02} \\
	&   	     & {\it et al.} \cite{sriganesh00} & Scheu \cite{kroger98} & \\
\hline
0.125 & 1.23(13) & 1.22(2) & 1.314 & 1.220 \\
0.25  & 1.24(17) & 1.24(3) & 1.279 & 1.230 \\
0.5   & 1.21(19) & 1.20(3) & 1.227 & 1.165 \\
1     & 1.12(17) & 1.12(3) & 1.128 & 0.99 \\
2     & 0.99(13) & 1.00(2) & 0.991 & 0.88 \\
4     & 0.84(8)  & 0.85(2) & 0.837 & 1.07 \\
8     & 0.69(5)  & 0.68(1) & 0.690 & 1.78  \\
16    & 0.56(3)  & 0.56(1) & 0.559 & 3.41  \\
\end{tabular}
\end{table}
\begin{table}
\caption{Results of numerical integration over symmetric and antisymmetric states for two operators.
\label{tab4} }
\begin{center}
\begin{tabular}{ccc}
Integral &  Symmetric & Antisymmetric  \\
\hline
$ \langle \frac{\partial^4}{\partial z^4} \rangle_0 $ & -0.577655 &  1.093349\\
$ \langle \delta(z) \rangle_0 $ & 0.490777 & 0.0\\
\end{tabular}
\end{center}
\end{table}
\center
\widetext
\input psfig
\psfull

\end{document}